\documentclass[a4paper,11pt]{article}
\pdfoutput=1
\usepackage[a4paper, margin=1in]{geometry}
\usepackage{amssymb,amsmath,amsfonts}
\usepackage[normalem]{ulem}
\usepackage[utf8x]{inputenc}
\usepackage{slashed}
\usepackage{graphicx}
\usepackage{tabularx}
\usepackage{here}
\usepackage{color}
\usepackage{csquotes} 
\usepackage{comment}
\usepackage{mathrsfs}
\usepackage{float}
\usepackage{ascmac}
\usepackage{multirow}
\usepackage{longtable}
\usepackage{bm}
\usepackage{ulem}
\usepackage[italicdiff]{physics}
\usepackage{booktabs}
\usepackage{appendix}

\makeatletter
\newcommand*\rel@kern[1]{\kern#1\dimexpr\macc@kerna}
\newcommand*\widebar[1]{%
  \begingroup
  \def\mathaccent##1##2{%
    \rel@kern{0.8}%
    \overline{\rel@kern{-0.8}\macc@nucleus\rel@kern{0.2}}%
    \rel@kern{-0.2}%
  }%
  \macc@depth\@ne
  \let\math@bgroup\@empty \let\math@egroup\macc@set@skewchar
  \mathsurround\z@ \frozen@everymath{\mathgroup\macc@group\relax}%
  \macc@set@skewchar\relax
  \let\mathaccentV\macc@nested@a
  \macc@nested@a\relax111{#1}%
  \endgroup
}
\makeatother

\numberwithin{equation}{section}

\def\bea{\begin{eqnarray}}
\def\eea{\end{eqnarray}}

\title{
Probing the Geometry of Viable Froggatt–Nielsen-like Flavor Textures
}
\author{Davide Meloni\footnote{davide.meloni@uniroma3.it}\\
        \small $^{1}$Department of Mathematics and Physics, Roma Tre University, Rome (Italy)
}
\begin{document}

\maketitle
\begin{abstract} 
We investigate the geometry of viable Froggatt--Nielsen-like flavor textures in a controlled toy landscape. In this first texture-level analysis, each point is represented by the eighteen integer exponents entering the up- and down-type Yukawa matrices, without imposing the additional charge-factorization constraints of specific Froggatt--Nielsen models. Viability is imposed through cuts on quark mass ratios and selected CKM observables.
We construct four benchmark ensembles, ranging from masses-only constraints to the inclusion of the full set of CKM observables considered in this work, and study the resulting point clouds in the raw exponent space. Principal-component analysis shows no evidence for strong linear compression: in all benchmarks, fifteen principal components are required to account for 90\% of the variance. Nearest-neighbour intrinsic-dimensionality diagnostics, including TwoNN and Levina--Bickel estimators, support the same qualitative conclusion, yielding high effective dimensions of order \(10\)--\(12\).
These results do not rule out hidden flavor-theory geometry in other coordinates; rather, they suggest that raw Froggatt--Nielsen exponents may not provide natural coordinates for revealing such a geometry.
\end{abstract}

\section{Introduction}

The origin of the observed pattern of fermion masses and flavor mixing remains one of the central unresolved problems in particle physics. While the Standard Model successfully parametrizes quark and lepton masses through Yukawa couplings, it provides no explanation for the hierarchies observed in the fermion sector. The enormous spread of quark masses across several orders of magnitude, together with the highly structured pattern of flavor mixing encoded in the Cabibbo--Kobayashi--Maskawa (CKM) matrix~\cite{Cabibbo:1963yz,Kobayashi:1973fv}, strongly suggests the existence of organizing principles beyond the Standard Model description.

Over the past decades, a large variety of flavor models have been proposed to address this problem. Among the most influential frameworks are Froggatt--Nielsen (FN) constructions \cite{Froggatt:1978nt}, in which hierarchical Yukawa couplings emerge from powers of a small expansion parameter associated with a spontaneously broken flavor symmetry. Variants of this idea have generated an extensive model-building literature, including phenomenological classifications of viable Yukawa textures \cite{Fritzsch:1977vd,Frampton:1985wz,
Ramond:1993kv,Fritzsch:2002ga,Xing:2003yj,Ponce:2011va,Ludl:2015lta},
gauged anomaly-free or anomaly-controlled realizations \cite{Rathsman:2019qkq,Smolkovic:2019jow,Bonnefoy:2019lsn}, systematic numerical scans and optimization-based searches
\cite{Fedele:2020fvh,Cornella:2023zme,Ibe:2024cvi} and ultraviolet- or string-inspired constructions 
\cite{Dudas:2009hu,King:2010uy,Leontaris:2010zd,Leontaris:2009pt,Constantin:2024yxh}.

A remarkable feature of this landscape is its apparent redundancy. Systematic studies of FN charge assignments and texture zeros have emphasized that many distinct charge choices or matrix structures can reproduce the observed quark masses and mixings, rather than singling out a unique preferred texture.
Concrete flavor-symmetry models, including examples involving not only quarks but also leptons, illustrate the same point: phenomenologically successful flavor structures can often be obtained from many different assignments of charges, representations, or texture parameters~\cite{Altarelli:2010gt,King:2013eh}. Consequently, the flavor problem may not be solely a question of identifying a single successful model, but also of understanding the global organization of the space of viable theories itself.

Recent work has begun to address this issue using algorithmic and computational methods. Reinforcement-learning searches have shown that automated exploration can identify viable quark mass models in high-dimensional model-building spaces~\cite{Harvey:2021oue,Nishimura:2020nre} (see also \cite{Giarnetti:2025mit} for a recent review), while recent FN scans and Bayesian/systematic studies explicitly map large sets of viable charge assignments rather than focusing on isolated benchmark textures~\cite{Fedele:2020fvh,Cornella:2023zme,Constantin:2024yxh,Cornella:2025lff,Ibe:2024cvi}.

This observation motivates a broader perspective. Rather than studying individual flavor models in isolation, one may regard flavor theories as points in a high-dimensional theory space and ask whether the subset compatible with experimental observations exhibits an underlying geometric structure. If viable theories populate theory space in a highly organized manner, one might expect them to concentrate around lower-dimensional manifolds or other compressed structures. Conversely, if viable theories remain broadly distributed throughout parameter space, the effective dimensionality of the landscape would remain large.
This motivates a geometric viewpoint on the flavor problem: rather than asking
only which individual models reproduce the data, one may ask how the set of
viable flavor theories is organized as a whole. In such a picture, apparently
distinct flavor models could correspond to nearby points in an underlying
representation, and the observed flavor hierarchies might reflect deeper
organizing principles. Determining whether such a structure exists requires
quantitative probes of the geometry of viable flavor theories.

The purpose of the present work is to perform an exploratory investigation of this question within a controlled toy-model setting. We consider ensembles of viable Froggatt--Nielsen-like flavor textures and study their distribution in the space of FN exponents. Rather than imposing the charge-factorization structure of specific Froggatt--Nielsen models, we treat the eighteen exponents as independent texture coordinates. This defines a broad texture-level landscape and provides a useful baseline in which to test whether phenomenological viability alone induces geometric organization or dimensional compression.

To address this question we employ two complementary approaches. First, we
perform a principal-component analysis (PCA) of the accepted texture ensembles.
PCA is a standard dimensional-reduction technique, widely used to identify the
dominant directions of variation in high-dimensional datasets, with applications
in high-energy and astroparticle physics
\cite{Hotelling:1933,Jolliffe:2002pca,
Huterer:2002hy,Bhalerao:2014mua,Bodwin:2019ivc}.
In the present context, it probes linear compression and provides a simple
diagnostic of whether the viable landscape is concentrated near a
low-dimensional linear subspace. Second, we apply nearest-neighbour intrinsic-dimensionality estimators, in
particular the TwoNN method of Ref.~\cite{Facco:2017} and the
Levina--Bickel estimator~\cite{Levina:2004}. These methods probe the local
scaling of distances within the accepted point cloud and are therefore
sensitive to possible nonlinear manifold structure that would not be captured
by PCA.
Together,
these two diagnostics allow us to distinguish linear compression from a more
general reduction of effective dimensionality.

Our analysis is based on distinct benchmark ensembles obtained by imposing progressively stronger phenomenological constraints, beginning with quark mass ratios alone and subsequently adding CKM observables. This hierarchy of benchmarks allows us to study how increasing flavor information affects the structure of the viable landscape.

The main result of the study is negative in a precise and informative sense. Across all benchmark datasets, PCA requires approximately fifteen principal components to account for 90\% of the variance in the full eighteen-dimensional FN exponent space. Independent nearest-neighbour intrinsic-dimensionality estimators  indicate effective dimensions of order $10$--$12$ under the primary analysis convention. While CKM constraints substantially reduce the volume of viable parameter space (as obviously expected) and increase the degree of organization visible in the leading principal components, they do not produce strong dimensional reduction.

We want to remark here that this result should not be interpreted as evidence against the existence of an underlying geometry of flavor theory space. Rather, it suggests that such a geometry is not strongly exposed in the raw FN exponent coordinates used in the present analysis. One possible interpretation is that Froggatt--Nielsen exponents may not be natural coordinates for revealing the geometry, and that a more compressed geometric structure may emerge only after an appropriate change of variables, metric, or representation.
The present work should therefore be viewed as a first step toward a broader program aimed at reconstructing and characterizing the geometry of viable flavor theories. Our goal is not to establish the absence of geometric organization, but rather to determine what can be learned from the simplest available coordinates and to identify the limitations of that description.

The paper is organized as follows. In Section~\ref{sec:model} we introduce the
toy Froggatt--Nielsen framework used in the analysis, define the exponent-space
representation of the textures, and describe the phenomenological selection
criteria used to construct the benchmark datasets. In Section~\ref{sec:pca} we
perform a principal-component analysis of the accepted texture ensembles and
use it to quantify their degree of linear compressibility. In
Section~\ref{sec:id} we complement this analysis with nearest-neighbour
intrinsic-dimensionality estimators, including the TwoNN method and the
Levina--Bickel estimator, in order to test for possible nonlinear
low-dimensional structure. We also discuss robustness checks associated with
finite-sample effects, discreteness, and metric dependence. Finally, in
Section~\ref{sec:conclusions} we summarize the results.

\section{Toy Model and Selection Procedure}
\label{sec:model}
In this section we define the simplified Froggatt--Nielsen setup used
throughout the analysis and describe how the benchmark ensembles of viable
textures are constructed. The purpose of the model is to generate a controlled and
interpretable set of Yukawa textures whose geometry can be studied in exponent
space. We first specify the texture parametrization and the construction of
quark masses and mixings, and then introduce the acceptance criteria used to
select phenomenologically viable points.
\subsection{Froggatt--Nielsen texture parametrization}

We consider a simplified Froggatt--Nielsen-inspired framework in which the up- and down-type Yukawa matrices are parametrized as
\begin{equation}
(Y_u)_{ij}
=
a^{(u)}_{ij}\,
\epsilon^{\,n^{(u)}_{ij}},
\qquad
(Y_d)_{ij}
=
a^{(d)}_{ij}\,
\epsilon^{\,n^{(d)}_{ij}},
\end{equation}
with
\begin{equation}
\epsilon = 0.22.
\end{equation}
This parametrization follows the standard FN logic that flavor hierarchies are encoded in powers of a small symmetry-breaking parameter, while leaving the unknown order-one coefficients as local, non-hierarchical data.
The coefficients $a^{(u,d)}_{ij}$ are taken to be real $\mathcal{O}(1)$ numbers, while the exponents
\begin{equation}
n^{(u,d)}_{ij}\ge 0
\end{equation}
are restricted to non-negative integers.
Each texture is therefore characterized by eighteen integer-valued FN exponents,
\begin{equation}
\mathbf{x}
=
(n^{(u)}_{11},\ldots,n^{(u)}_{33},
 n^{(d)}_{11},\ldots,n^{(d)}_{33}),
\end{equation}
which define an 18-dimensional exponent space. Throughout this work, these exponent vectors are used as a convenient coordinate representation of the flavor landscape. A central goal of our analysis is to assess whether the geometric structure of viable textures is naturally captured in these coordinates, or whether a more appropriate description may exist in terms of alternative variables.

\subsection{Masses and mixing}
In this exploratory study we work with real Yukawa coefficients
\(a^{(u,d)}_{ij}\). The Yukawa matrices are therefore real, the matrices
\(U_u\) and \(U_d\) may be taken orthogonal, and the CKM matrix obtained in
this setup has no physical CP-violating phase. Accordingly, we impose only
quark mass ratios and CKM moduli, and do not include the CKM phase or the
Jarlskog invariant among the selection observables. This choice is motivated
by the goal of the present work: to isolate the geometry of the FN exponent
landscape associated with flavor hierarchies and mixing magnitudes. The
description of CP violation requires additional information about the complex
phases of the order-one coefficients, and possibly about their ultraviolet
origin. Such information is not encoded in the integer exponent vectors alone
and will be considered in future extensions of this analysis.
The quark-sector mass matrices are thus constructed as
\begin{equation}
M_u^2 = Y_uY_u^T,
\qquad
M_d^2 = Y_dY_d^T.
\end{equation}
They are diagonalized according to
\begin{equation}
U_u^T M_u^2 U_u
=
{\rm diag}
(m_u^2,m_c^2,m_t^2),
\end{equation}

\begin{equation}
U_d^T M_d^2 U_d
=
{\rm diag}
(m_d^2,m_s^2,m_b^2),
\end{equation}
with eigenvalues ordered increasingly. The CKM matrix~\cite{Cabibbo:1963yz,Kobayashi:1973fv,ParticleDataGroup:2024cfk} is then defined as
\begin{equation}
V_{\rm CKM}
=
U_u^T U_d.
\end{equation}
The target observables used in the analysis are
\begin{align}
\frac{m_u}{m_c} &= 2.0\times10^{-3},
&
\frac{m_c}{m_t} &= 2.678\times10^{-3},
\\
\frac{m_d}{m_s} &= 5.0\times10^{-2},
&
\frac{m_s}{m_b} &= 1.370\times10^{-2},
\end{align}
together with
\begin{eqnarray}
|V_{us}| = 0.2245, \qquad 
|V_{cb}| = 0.041, \qquad
|V_{ub}| = 0.00382.
\end{eqnarray}

The numerical target values are chosen as representative quark-mass and CKM
hierarchy inputs, guided by current flavor data summaries and by standard
running-mass determinations. Since the selection uses broad multiplicative
windows, these targets should not be interpreted as a precision fit to quark
masses at a fixed renormalization scale~\cite{ParticleDataGroup:2024cfk,Huang:2020hdv}.

\subsection{Acceptance criteria}
A texture is classified as viable if all selected observables satisfy independent multiplicative bounds relative to their target values. For a generic observable $O$, we define the multiplicative deviation
\begin{equation}
f(O)
=
\max
\left(
\frac{O_{\rm model}}{O_{\rm target}},
\frac{O_{\rm target}}{O_{\rm model}}
\right).
\end{equation}
The mass and CKM cuts are chosen with different tolerances because they play
different roles in the FN texture construction. For quark mass ratios we use a
relatively broad window, \(f(O)\leq 5\). This is appropriate for an
order-of-magnitude FN analysis: a change by one unit in an exponent corresponds
to a factor \(1/\epsilon\simeq 4.5\), and the unknown order-one coefficients
can easily shift mass ratios by comparable factors. The mass cuts should
therefore be interpreted as selecting textures with the correct hierarchical
orders rather than as performing a precision fit. For CKM observables we impose
a tighter condition, \(f(O)\leq 1.5\), because the mixing angles are accurately
measured and because the purpose of the CKM benchmarks is to test how
additional flavor information reshapes the viable exponent landscape. We outline that these
cuts are not statistical confidence intervals but define, instead, phenomenological
acceptance windows suitable for constructing controlled benchmark ensembles.
Acceptance is determined by these independent per-observable cuts. No global goodness-of-fit quantity is used in the selection procedure.
To avoid overcounting, accepted textures with identical 18-dimensional exponent vectors are removed. 
Each benchmark dataset therefore consists of unique points in FN exponent space.
We construct four benchmark datasets, each containing 1000 accepted textures\footnote{The sample size was chosen for practical computational reasons. In particular, the acceptance rate decreases as CKM constraints are added, making
larger benchmark ensembles substantially more expensive to generate. The
resulting datasets are intended for exploratory geometric diagnostics rather
than for a precision statistical characterization of the full viable
landscape.}, by imposing a progressively richer set of flavor constraints. The first benchmark is selected using only quark mass ratios, and therefore probes the geometric structure induced by the observed fermion mass hierarchy alone. We then add the CKM observables step by step, starting from the Cabibbo angle $|V_{us}|$, then including $|V_{cb}|$, and finally imposing the full set of mixing constraints considered in this work, namely $|V_{us}|$, $|V_{cb}|$, and $|V_{ub}|$.
This sequence is designed to disentangle the geometric impact of different sectors of flavor data. In particular, it allows us to ask whether the quark mass hierarchy by itself already selects a geometrically organized region of FN exponent space, or whether the observed mixing structure is essential for producing additional organization. Comparing the four benchmarks therefore provides a controlled way to track how the viable landscape changes as increasingly restrictive and phenomenologically informative constraints are imposed.

\section{Principal-Component Analysis }
\label{sec:pca}

As a first probe of geometric organization, we perform a principal-component
analysis (PCA) of the accepted FN exponent vectors
\cite{Jolliffe:2002pca}. 
PCA is a standard tool for diagnosing linear variance structure and possible low-dimensional compression in high-dimensional data~\cite{Hotelling:1933,Jolliffe:2002pca,Bishop:2006prml}. It provides a first, model-independent diagnostic of the large-scale geometry of the accepted FN textures in the 18-dimensional exponent space. Each accepted texture is represented by a point
\[
\mathbf{x}
=
(n^{(u)}_{11},\ldots,n^{(u)}_{33},
 n^{(d)}_{11},\ldots,n^{(d)}_{33}) ,
\]
and the full set of accepted textures defines a point cloud in this space. PCA analyzes the shape of this point cloud by identifying the directions along which it is most widely spread.
More precisely, PCA rotates the original coordinate system to a new set of mutually orthogonal axes, called principal components. These axes are ordered according to the amount of variance that they explain. In this context, variance means the spread of the accepted textures in exponent space: a direction has large variance if the projections of the accepted textures along that direction vary significantly from one texture to another, and small variance if the textures are nearly concentrated at similar values along that direction. The total variance of the dataset is therefore a measure of the overall dispersion of the accepted point cloud, while the variance explained by a given principal component measures the fraction of this total dispersion captured along that direction.

The {\it first} principal component is the direction of maximal spread of the accepted textures. The {\it second} principal component is the orthogonal direction with the largest remaining spread, and so on. The resulting PCA spectrum quantifies how the total variance is distributed among these orthogonal directions.

This provides a useful measure of linear compressibility. If the viable textures were concentrated close to a low-dimensional linear subspace of the full 18-dimensional exponent space, most of the total variance would be captured by only a small number of principal components. Conversely, if many principal components are required to account for a large fraction of the variance, this indicates that the viable textures remain broadly distributed across the ambient exponent space. We therefore use the PCA spectrum as a quantitative diagnostic of whether the flavor constraints select a linearly compressed geometric structure in FN exponent space. In particular, we monitor both the variance explained by the leading principal components and the number of components required to reach fixed cumulative variance thresholds.

To quantify this behavior we report the fractions of variance explained by the leading principal components and define
\begin{equation}
N_{50},
\qquad
N_{75},
\qquad
N_{90},
\end{equation}
as the numbers of principal components required to explain 50\%, 75\%, and 90\% of the total variance, respectively.

\subsection{Results}

The PCA results for the four benchmark datasets are summarized in
Table~\ref{tab:pca_results}. For each dataset we report the fraction of
total variance explained by the first two principal components, together
with the number of components required to reach 50\%, 75\%, and 90\% of
the cumulative explained variance, denoted by $N_{50}$, $N_{75}$, and
$N_{90}$, respectively.

Table~\ref{tab:pca_results} already illustrates the two main trends emerging
from the PCA analysis. First, the leading principal components become
progressively more important as additional CKM observables are imposed. The
fraction of variance explained by the first principal component increases from
9.89\% in the masses-only benchmark to 15.16\% when the full set of CKM
constraints is included, while the second principal component follows a similar
trend. This indicates that flavor-mixing information makes the accepted
distribution increasingly anisotropic in exponent space.
At the same time, however, the global dimensionality of the accepted landscape
changes very little. Although the leading principal components become more
prominent, the number of components required to reproduce most of the variance
remains remarkably stable. In particular, all four benchmark datasets satisfy
\(N_{90}=15\), showing that the viable FN textures continue to occupy a large
fraction of the original eighteen-dimensional exponent space. Thus, the CKM
constraints increase the organization of the landscape without producing a
strong linear compression.

\begin{table}[t]
\centering
\begin{tabular}{lccccc}
\hline
Benchmark & PC1 (\%) & PC2 (\%) & $N_{50}$ & $N_{75}$ & $N_{90}$ \\
\hline
Masses only
& 9.89
& 9.33
& 7
& 12
& 15
\\

Masses + $|V_{us}|$
& 10.13
& 9.52
& 6
& 11
& 15
\\

Masses + $|V_{us}|+|V_{cb}|$
& 10.57
& 9.99
& 6
& 11
& 15
\\

Masses + $|V_{us}|+|V_{cb}|+|V_{ub}|$
& 15.16
& 14.13
& 5
& 11
& 15
\\
\hline
\end{tabular}
\caption{\it 
Principal-component analysis of the four benchmark datasets. PC1 and PC2
denote the fractions of total variance explained by the first and second
principal components. The quantities $N_{50}$, $N_{75}$, and $N_{90}$
denote the number of principal components required to reach 50\%, 75\%,
and 90\% cumulative explained variance.
}
\label{tab:pca_results}
\end{table}

\begin{figure}[t]
\centering
\includegraphics[width=0.78\textwidth]{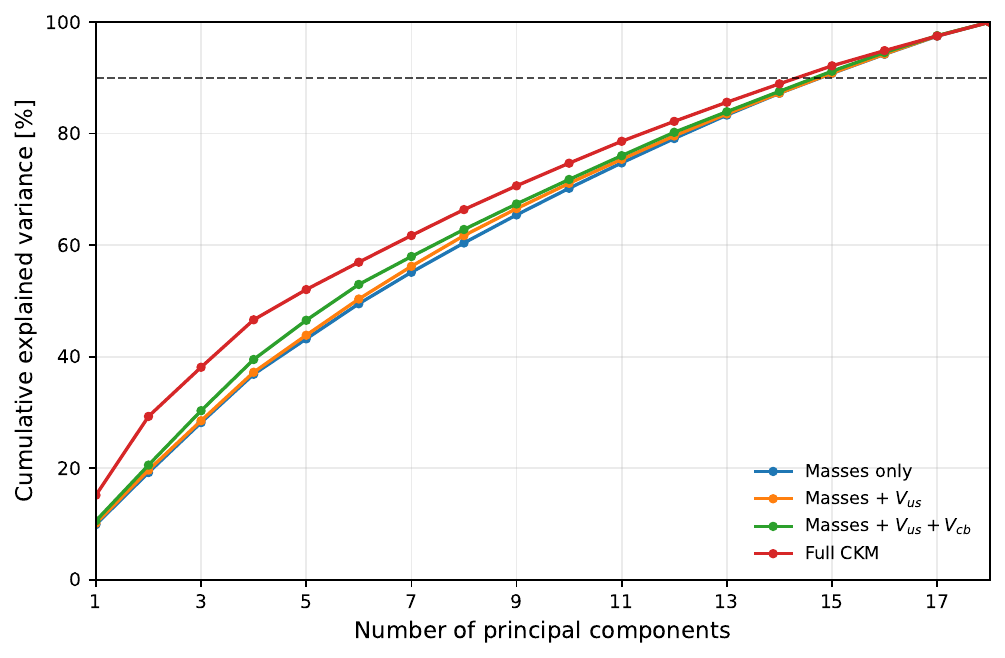}
\caption{\it 
Cumulative explained variance as a function of the number of principal
components for the four benchmark datasets. The dashed horizontal line
indicates the 90\% cumulative-variance threshold. The full-CKM benchmark
displays a larger fraction of variance in the leading components, showing
that the inclusion of mixing observables increases the anisotropy of the
accepted point cloud. However, all four benchmarks require fifteen
principal components to reach the 90\% threshold, indicating that the
viable textures are not compressed into a low-dimensional linear subspace
of the 18-dimensional FN exponent space.
}
\label{fig:pca_cumulative_variance}
\end{figure}

\begin{figure}[t]
\centering
\includegraphics[width=0.78\textwidth]{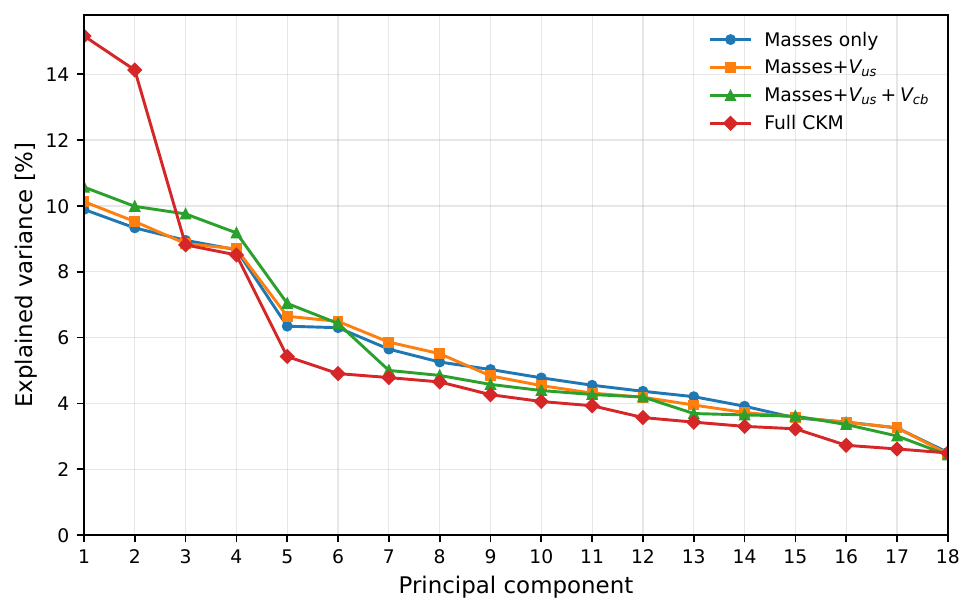}
\caption{\it 
Explained variance ratio of the individual principal components for the
four benchmark datasets. Each point shows the fraction of the total
variance carried by a single principal component. The full-CKM benchmark
shows a clear enhancement of the first two components, indicating that
mixing constraints make the accepted point cloud more anisotropic.
Nevertheless, the spectra remain broad and exhibit a long tail extending
to high component number. Thus, even when the leading directions become
more pronounced, the variance is still distributed over many independent
directions in the 18-dimensional exponent space.
}
\label{fig:pca_spectra}
\end{figure}

The cumulative PCA spectra are shown in
Fig.~\ref{fig:pca_cumulative_variance}, while the corresponding individual
PCA spectra are shown in Fig.~\ref{fig:pca_spectra}. Together, these two
figures provide complementary views of the same geometric information.
Fig.~\ref{fig:pca_cumulative_variance} shows how rapidly this variance is
accumulated as more components are retained, whereas Fig.~\ref{fig:pca_spectra} shows how the total variance is distributed
among individual principal components.

The most important feature of Fig.~\ref{fig:pca_cumulative_variance} is
the slow rise of the cumulative explained variance. In all four benchmark
datasets, the cumulative curve increases gradually rather than saturating
after only a few principal components. Since, as commented above, $N_{90}=15$ for every dataset considered, 
the  90\% of the variance is captured only
after retaining most of the available principal directions. The
masses-only benchmark already exhibits this behavior: even when only the
quark mass hierarchy is imposed, fifteen principal components are required
to account for 90\% of the total spread of the accepted point cloud.

The individual spectra in Fig.~\ref{fig:pca_spectra} confirms this
interpretation. In the masses-only benchmark, the first principal
component accounts for only 9.89\% of the variance, and the second for
9.33\%. The subsequent components decrease gradually, rather than showing
a sharp drop after one or two dominant directions. 
The inclusion of CKM information modifies this picture in a visible but
controlled way. As mixing constraints are added, the leading principal
components become more important. The variance explained by the first
principal component increases from 9.89\% in the masses-only benchmark to
10.13\% after imposing $|V_{us}|$, to 10.57\% after also imposing
$|V_{cb}|$, and finally to 15.16\% in the full-CKM benchmark. The second
principal component follows the same trend, rising from 9.33\% to
14.13\%. 
However, the same Fig.~\ref{fig:pca_spectra} also shows why this effect should not be
interpreted as strong dimensional reduction. Although the full-CKM
benchmark is more strongly dominated by its leading components, the
spectrum does not collapse after the first few directions. Instead, a
long tail of subleading components remains present across the full range
of principal-component indices. This explains why, even in the most
constrained benchmark, five components are still required to explain 50\%
of the variance, eleven are required to explain 75\%, and fifteen are
required to explain 90\%.

The PCA analysis therefore reveals two simultaneous effects. On the one
hand, the CKM constraints introduce additional geometric organization in
the viable region of FN exponent space, making the accepted point cloud
more anisotropic and increasing the weight of the leading principal
directions. On the other hand, this organization remains distributed over
many directions and does not amount to a collapse onto a low-dimensional
linear manifold. The viable textures continue to populate a high-dimensional
region when viewed in the raw 18-dimensional exponent coordinates.

The conservative conclusion is that the observed quark masses and mixings
do impose nontrivial geometric structure on the FN landscape, but this
structure is not well described as strong linear compression in the
original exponent variables. This motivates the complementary nonlinear
dimensionality analysis performed below, and also leaves open the
possibility that a more compressed description may require coordinates
different from the raw FN exponents.

\section{Intrinsic-Dimensionality Analysis}
\label{sec:id}

While PCA provides a useful probe of linear compression, it is insensitive to nonlinear manifold structure. A dataset may require many principal components to reproduce its variance and yet still lie on a lower-dimensional curved manifold embedded in a higher-dimensional ambient space. To investigate this possibility, we complement the PCA study with nearest-neighbour intrinsic-dimensionality estimators.

Intrinsic-dimensionality estimators provide a complementary nonlinear
diagnostic. Their numerical estimates can nevertheless depend on sampling
density, sample size, noise, boundary effects, and the neighbourhood scale
used in the analysis~\cite{Verveer:1995,Theiler:1990,Eckmann:1992,
Campadelli:2015,Camastra:2016}. For nearest-neighbour estimators, the choice
of distance metric and violations of the locally continuous-distribution
assumption require additional care, particularly for discrete,
lattice-valued datasets with degenerate nearest-neighbour
distances~\cite{Levina:2004,Facco:2017,Denti:2022}.


\subsection{Intrinsic dimensionality from nearest-neighbour geometry}

Our primary estimator is the Two Nearest Neighbours (TwoNN) method~\cite{Facco:2017}, which belongs to the broader family of nearest-neighbour and scaling-based intrinsic-dimensionality diagnostics~\cite{Grassberger:1983zz,Camastra:2002,Levina:2004}.

The basic idea of TwoNN is that nearest-neighbour distances encode how rapidly
the number of data points grows with distance around a typical point. In a
low-dimensional distribution, the volume enclosed within a radius \(r\) grows
relatively slowly with \(r\). In a high-dimensional distribution, the same
volume grows much more rapidly. The relative distances to the first and second
nearest neighbours therefore contain information about the effective dimension
sampled by the point cloud.

For each accepted texture \(\mathbf{x}_i\), we compute the distances to its
first and second nearest neighbours,
\begin{equation}
r_{1,i},
\qquad
r_{2,i},
\end{equation}
and form the ratio
\begin{equation}
\mu_i = \frac{r_{2,i}}{r_{1,i}} .
\end{equation}
By construction, \(\mu_i \geq 1\). If the data locally sample a smooth
\(d\)-dimensional distribution with approximately uniform density, the
cumulative distribution of \(\mu\) is
\begin{equation}
F(\mu)
=
1-\mu^{-d},
\qquad
\mu \geq 1 \,,
\end{equation}
or, equivalently,
\begin{equation}
-\log\!\left[1-F(\mu)\right]
=
d \log \mu .
\label{equat}
\end{equation}
Thus, in the ideal TwoNN setting, the intrinsic dimension \(d\) is obtained as
the slope of a straight line in the plane
\begin{equation}
\left(
\log \mu,
-\log[1-F(\mu)]
\right).
\end{equation}

In practice, after removing the degenerate ratios with \(\mu_i=1\)\footnote{The abundance of these points is
recorded separately and used as a diagnostic of discreteness effects.}, we retain
\(N_{\rm eff}\) ratios and arrange them in increasing order,
\begin{equation}
\mu_{(1)} \leq \mu_{(2)} \leq \cdots
\leq \mu_{(N_{\rm eff})}.
\end{equation}
sample. We associate with the \(j\)-th ordered ratio the empirical cumulative
probability
\begin{equation}
F_j = \frac{j}{N_{\rm eff}+1}.
\end{equation}
The use of \(N_{\rm eff}+1\) ensures that \(F_j<1\) for every point, so that
\(-\log(1-F_j)\) remains finite.

For each ordered ratio, we then define
\begin{equation}
X_j=\log\mu_{(j)},
\qquad
Y_j=-\log(1-F_j).
\end{equation}
According to the ideal TwoNN relation, these quantities satisfy
\begin{equation}
Y_j=d\,X_j.
\end{equation}
We therefore estimate the intrinsic dimension \(d\) as the slope of a linear
regression through the origin in the \((X_j,Y_j)\) plane. To reduce the
influence of the distribution tails, the regression is restricted to points
between the 5th and 95th percentiles.
All nearest-neighbour distances in the primary analysis are computed using the
Euclidean metric in the 18-dimensional FN exponent space. Let
\(\mathbf{x}\) and \(\mathbf{y}\) denote two accepted textures, represented by
their exponent vectors,
\[
\mathbf{x}=(x_1,\ldots,x_{18}),\qquad
\mathbf{y}=(y_1,\ldots,y_{18}),
\]
where \(x_i\) and \(y_i\) are the FN exponents corresponding to the \(i\)-th
entry of the exponent vector. Their Euclidean distance is defined as
\begin{equation}
d_{L_2}(\mathbf{x},\mathbf{y})
=
\left(
\sum_{i=1}^{18}
(x_i-y_i)^2
\right)^{1/2}.
\label{eqdist}
\end{equation}

\begin{figure}[t]
\centering
\includegraphics[width=0.78\textwidth]{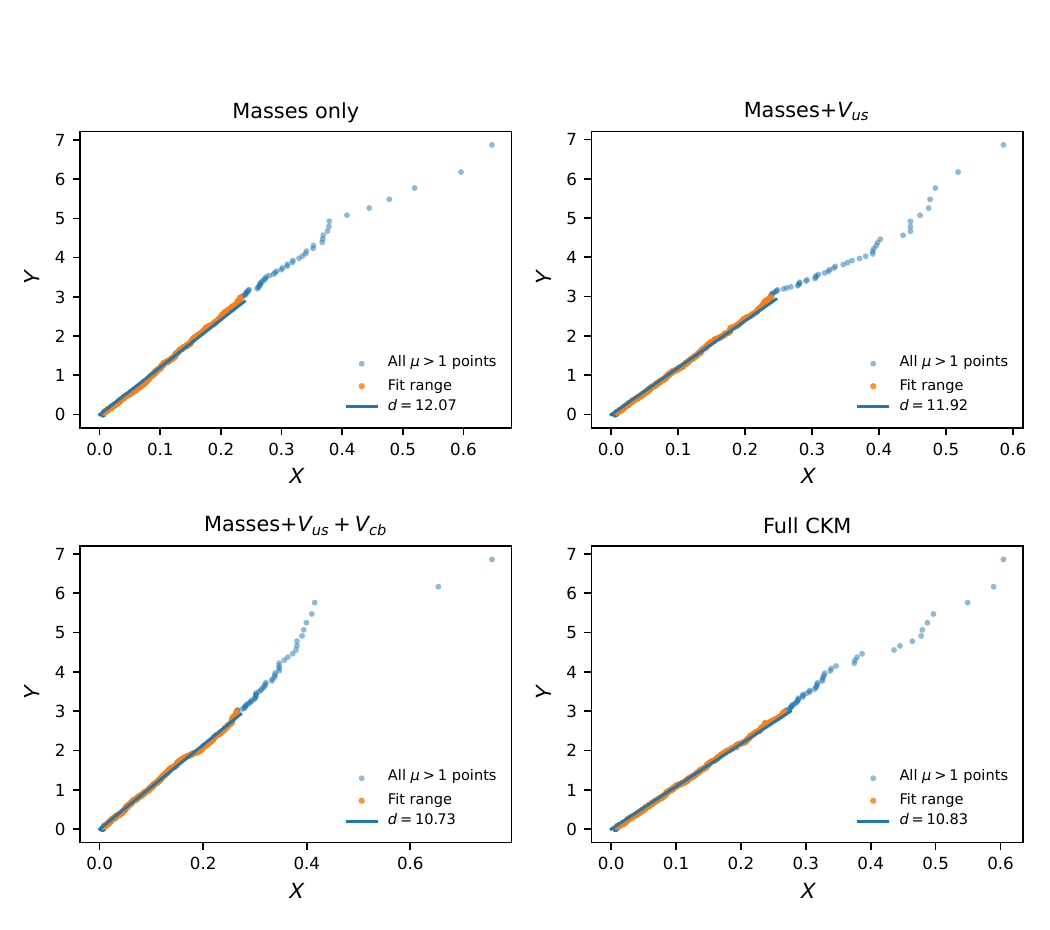}
\caption{\it 
Linearized TwoNN distributions for the four benchmark datasets. The horizontal
axis shows \(\log\mu\), while the vertical axis shows
\(-\log[1-F(\mu)]\). In the ideal TwoNN limit, the slope of each curve gives
the intrinsic dimension. The fits are performed through the origin over the
central 5--95\% quantile range after removing exactly degenerate ratios with
\(\mu=1\). 
}
\label{fig:twonn_linear_fits}
\end{figure}

Figure~\ref{fig:twonn_linear_fits} shows the corresponding linearized TwoNN
distributions for the four benchmark datasets. This plot gives a direct visual
representation of the fit used to extract the TwoNN dimension: the steeper the
curve, the larger the estimated intrinsic dimension. 
In addition, the resulting primary TwoNN estimates are summarized in
Table~\ref{tab:twonn_results}.
\begin{table}[t]
\centering
\begin{tabular}{lcccc}
\hline
Benchmark
& $N$
& $N(\mu=1)$
& $N_{\rm eff}$
& $d_{\rm TwoNN}$ \\
\hline
Masses only
& 1000
& 35
& 965
& 12.07
\\

Masses + $|V_{us}|$
& 1000
& 47
& 953
& 11.92
\\

Masses + $|V_{us}|+|V_{cb}|$
& 1000
& 47
& 953
& 10.73
\\

Masses + $|V_{us}|+|V_{cb}|+|V_{ub}|$
& 1000
& 53
& 947
& 10.83
\\
\hline
\end{tabular}
\caption{\it 
Primary TwoNN intrinsic-dimensionality estimates for the four benchmark
datasets, using Euclidean distances in the raw 18-dimensional FN exponent
coordinates. Points with exactly degenerate nearest-neighbour ratios
\(\mu=1\) are excluded from the fit. 
}
\label{tab:twonn_results}
\end{table}

Several observations from Fig.~\ref{fig:twonn_linear_fits} and
Table~\ref{tab:twonn_results} are in order. First, the estimated dimensions are consistently
large compared with the very-low-dimensional case. Across all four benchmarks,
the primary TwoNN estimates lie in the range
\begin{equation}
d_{\rm TwoNN} \simeq 10.7\text{--}12.1 .
\end{equation}
Thus, even allowing for the finite-sample and discreteness limitations of
nearest-neighbour estimators, the accepted textures do not resemble a point
cloud organized around smaller dimensional structure in the raw
FN exponent space.

Second, the inclusion of CKM constraints does not produce a dramatic collapse
of the intrinsic-dimensionality estimate. The masses-only benchmark gives
$d_{\rm TwoNN}=12.07$, whereas the full-CKM benchmark gives
$d_{\rm TwoNN}=10.83$.
The decrease is visible but moderate. This behaviour is consistent with the
PCA analysis: flavor-mixing constraints introduce additional geometric
organization, but they do not reduce the viable landscape to a sharply
low-dimensional structure.

Third, the number of exactly degenerate ratios \(\mu=1\) remains modest in the
primary raw-\(L_2\) analysis. It ranges from 35 to 53 points out of 1000,
corresponding to only a few percent of each benchmark dataset. This confirms
that discreteness effects are present, as expected for integer-valued FN
exponents, but they do not dominate the primary Euclidean TwoNN estimate.

The conservative interpretation is therefore that the viable FN textures have
a high effective dimensionality when probed by local nearest-neighbour geometry
in the raw exponent coordinates. The numerical values in
Table~\ref{tab:twonn_results} should not be interpreted as precise manifold
dimensions. Rather, they should be read as evidence against a strongly
compressed, very-low-dimensional description of the accepted textures in the
original 18-dimensional FN exponent space.

\subsection{Levina--Bickel nearest-neighbour cross-check}

To test whether the previous conclusions depend on this specific ratio statistic, we
perform an independent cross-check using the maximum-likelihood estimator
introduced by Levina and Bickel~\cite{Levina:2004} 
(see also \cite{Camastra:2002,Facco:2017}  for related nearest-neighbour estimators).

This estimator infers the intrinsic dimension from the relative distances
between each data point and its first \(k\) nearest neighbours. The parameter
\(k\) controls the size of the neighbourhood being probed: small values of
\(k\) emphasize very local geometric structure, whereas larger values probe
the point cloud over a broader scale.
For each accepted texture \(\mathbf{x}_i\), let
\[
T_1(\mathbf{x}_i),\ldots,T_k(\mathbf{x}_i)
\]
denote the distances from \(\mathbf{x}_i\) to its first \(k\) nearest
neighbours, excluding the point itself. 
If the accepted textures locally sample a smooth \(d\)-dimensional manifold,
the number of points contained within a ball of radius \(r\) grows
approximately as
\[
N(r)\propto r^d.
\]
Consequently, the distances to the first, second, ..., \(k\)-th nearest
neighbours also obey a characteristic scaling determined by the intrinsic
dimension. The Levina--Bickel estimator exploits this scaling to infer \(d\)
from the observed nearest-neighbour distances.
This is given by:
\begin{equation}
\widehat d_k(\mathbf{x}_i)
=
\left[
\frac{1}{k-1}
\sum_{j=1}^{k-1}
\log
\frac{T_k(\mathbf{x}_i)}{T_j(\mathbf{x}_i)}
\right]^{-1}.
\end{equation}

In this paper we use the sample-averaged local Levina--Bickel estimator,
obtained by taking the arithmetic mean of the local estimates over all
accepted textures,
\begin{equation}
\widehat d_{\rm LB}(k)
=
\frac{1}{N}
\sum_{i=1}^{N}
\widehat d_k(\mathbf{x}_i).
\end{equation}
The analysis is performed using the same raw 18-dimensional FN exponent
coordinates and the same Euclidean metric as in the primary TwoNN study. We
consider
\[
k = 5,\ 10,\ 20,\ 30 ,
\]
which allows us to test the stability of the estimate as the neighbourhood
scale is varied. 
We do not interpret any single
choice of \(k\) as defining a preferred intrinsic dimension. Instead, the
dependence of \(\widehat d_{\rm LB}(k)\) on \(k\) is used as a robustness
diagnostic.

\begin{figure}[t]
\centering
\includegraphics[width=0.78\textwidth]{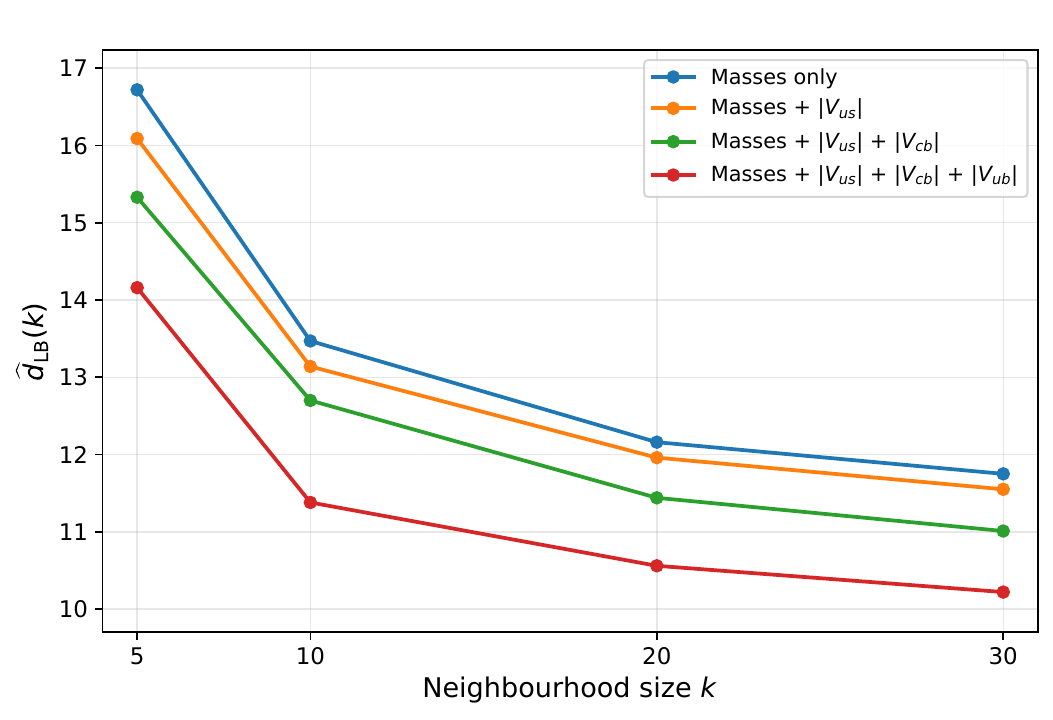}
\caption{
\it Levina--Bickel intrinsic-dimensionality estimates as a function of the
neighbourhood size \(k\) for the four benchmark datasets. The estimates
decrease as \(k\) is increased, reflecting the dependence of local
nearest-neighbour estimators on the scale over which the point cloud is
probed. 
}
\label{fig:lb_k_dependence}
\end{figure}

We remind that the four benchmark datasets contain \(N=1000\) accepted textures, with 1000
unique exponent vectors in each case. No duplicate exponent vectors and no
zero nearest-neighbour distances are found. The resulting Levina--Bickel
estimates are summarized in Table~\ref{tab:lb_results} and Fig.~\ref{fig:lb_k_dependence}.

\begin{table}[t]
\centering
\begin{tabular}{lcccc}
\hline
Benchmark
& $\widehat d_{\rm LB}(5)$
& $\widehat d_{\rm LB}(10)$
& $\widehat d_{\rm LB}(20)$
& $\widehat d_{\rm LB}(30)$ \\
\hline
Masses only
& 16.72
& 13.47
& 12.16
& 11.75
\\

Masses + $|V_{us}|$
& 16.09
& 13.14
& 11.96
& 11.55
\\

Masses + $|V_{us}|+|V_{cb}|$
& 15.33
& 12.70
& 11.44
& 11.01
\\

Masses + $|V_{us}|+|V_{cb}|+|V_{ub}|$
& 14.16
& 11.38
& 10.56
& 10.22
\\
\hline
\end{tabular}
\caption{
\it Levina--Bickel intrinsic-dimensionality estimates for the four benchmark
datasets, computed in the raw 18-dimensional FN exponent coordinates using
Euclidean distances. Several neighbourhood sizes \(k\) are shown in order to
test the dependence of the estimate on the local scale. The estimates decrease
as \(k\) increases, but remain in a high-dimensional regime for all benchmarks.
}
\label{tab:lb_results}
\end{table}

Some comments are in order. First, the Levina--Bickel estimates are
systematically high for all four benchmark datasets. Even at the largest
neighbourhood size considered, \(k=30\), the estimates lie between
\[
\widehat d_{\rm LB}(30) = 10.22
\]
for the full-CKM benchmark and
\[
\widehat d_{\rm LB}(30) = 11.75
\]
for the masses-only benchmark. Thus, within this independent
nearest-neighbour estimator, once again the accepted textures do not resemble a
very-low-dimensional point cloud in the raw FN exponent coordinates.

Then, the estimates decrease as \(k\) is increased. This behaviour is not
unexpected: changing \(k\) changes the local scale over which the point cloud
is sampled, and nearest-neighbour dimensionality estimators are known to be
sensitive to finite-sample effects and local-density variations. For this
reason, we do not assign physical significance to the precise numerical value
obtained at any single \(k\). The relevant observation is that the estimates
remain high over the full range of neighbourhood sizes considered.

Finally, the ordering of the benchmark datasets is physically meaningful. The
full-CKM benchmark gives the smallest Levina--Bickel estimates for all values
of \(k\), while the masses-only benchmark gives the largest estimates. This is
consistent with the PCA and TwoNN analyses: adding mixing observables imposes
additional geometric organization on the accepted textures. However, the size
of the effect is moderate. The viable point cloud becomes more organized, but
it does not collapse onto a low-dimensional structure.

It is useful to compare the Levina--Bickel estimates directly with the primary
TwoNN results. This comparison is shown in Table~\ref{tab:twonn_lb_summary}.
\begin{table}[t]
\centering
\begin{tabular}{lccc}
\hline
Benchmark
& $d_{\rm TwoNN}$
& $\widehat d_{\rm LB}(10)$
& Range of $\widehat d_{\rm LB}(k)$ \\
\hline
Masses only
& 12.07
& 13.47
& 11.75--16.72
\\
Masses + $|V_{us}|$
& 11.92
& 13.14
& 11.55--16.09
\\
Masses + $|V_{us}|+|V_{cb}|$
& 10.73
& 12.70
& 11.01--15.33
\\
Masses + $|V_{us}|+|V_{cb}|+|V_{ub}|$
& 10.83
& 11.38
& 10.22--14.16
\\
\hline
\end{tabular}
\caption{\it 
Comparison between the primary TwoNN estimate, \(d_{\rm TwoNN}\), and the
Levina--Bickel estimate evaluated at \(k=10\),
\(\widehat d_{\rm LB}(10)\), for the four benchmark datasets. The last column
reports the range spanned by \(\widehat d_{\rm LB}(k)\) when
\(k=5,10,20,30\). Although the numerical estimates depend on the estimator and
on the neighbourhood size, both diagnostics consistently indicate that the
accepted textures remain high-dimensional in the raw FN exponent coordinates.
}
\label{tab:twonn_lb_summary}
\end{table}
The comparison shows that TwoNN and Levina--Bickel do not give identical
numbers, as expected for finite datasets in a discrete high-dimensional
exponent space. Nevertheless, they support the same qualitative conclusion.
Both estimators yield effective dimensions of order
\[
d_{\rm eff} \sim 10\text{--}12
\]
or higher, depending on the neighbourhood scale. In particular, neither
estimator gives values compatible with a sharply compressed  dimensional structure.

The Levina--Bickel cross-check therefore strengthens the interpretation of the
TwoNN analysis. The absence of strong low-dimensional compression is not an
artifact of using the nearest-neighbour ratio \(\mu=r_2/r_1\). Instead, it
persists when the local scaling of several nearest-neighbour distances is used.
Together with the PCA results, this supports a consistent picture: flavor
constraints impose nontrivial geometric organization on the viable FN
landscape, but in the raw 18-dimensional exponent coordinates this
organization remains broadly high-dimensional.

\subsection{Robustness of the intrinsic-dimensionality results}
\label{sec:id_robustness_summary}

We performed a series of checks to determine whether the nearest-neighbour
results could be driven by duplicate textures, degenerate distances, the
choice of metric, the TwoNN fitting convention, or the finite number of
available textures. In this subsection we describe the relevant definitions
and summarize the numerical results of these tests.

For each point \(\mathbf{x}_i\), let \(r_{1,i}\) and \(r_{2,i}\) denote the
distances to its first and second nearest neighbours. We distinguish exact
duplicates, identified by \(r_{1,i}=0\), from nonzero nearest-neighbour ties,
for which \(r_{2,i}=r_{1,i}>0\). We record the fractions of points satisfying
these two conditions as diagnostics of duplicate and lattice-induced distance
degeneracies. No duplicate exponent vectors are present in any of the four FN benchmarks.
Under the primary raw Euclidean metric in \ref{eqdist}, 
only \(3.5\)--\(5.3\%\) of the points have
\(r_{2,i}=r_{1,i}>0\), and hence \(\mu_i=r_{2,i}/r_{1,i}=1\).
These ratios contain no information about the slope of the linearized TwoNN
distribution and are therefore excluded from the fit. Their small abundance
shows that distance degeneracies are present, as expected for integer-valued
coordinates, but do not dominate the primary raw-\(L_2\) analysis.

We next test the sensitivity to the relative scale of the 18 exponent
coordinates. For this purpose, we introduce the standardized Euclidean
distance
\begin{equation}
d_{\rm std}(\mathbf{x},\mathbf{y})
=
\left[
\sum_{a=1}^{18}
\frac{(x_a-y_a)^2}{s_a^2}
\right]^{1/2},
\end{equation}
where \(s_a\) is the sample standard deviation of coordinate \(a\), evaluated
separately within each benchmark. This rescaling prevents coordinates with
larger sample variances from receiving a correspondingly larger weight in the
distance. The standardized TwoNN estimates lie between \(11.13\) and \(12.54\),
compared with \(10.73\)--\(12.07\) for the raw-\(L_2\) analysis. For a given
benchmark, the largest absolute change produced by standardization is \(0.47\).
The precise numerical estimates therefore depend moderately on the coordinate
scaling, but their high-dimensional character is unchanged.

The TwoNN estimate is obtained from a through-origin fit to the linearized
empirical distribution \ref{equat}, that we rewrite here as:
\begin{equation}
-\log\!\left[1-F(\mu)\right]
=
\widehat d_{\rm TwoNN}\log\mu.
\end{equation}
To test the dependence on the portion of the distribution retained in the fit,
we repeat the analysis using the central \(1\)--\(99\%\), \(5\)--\(95\%\),
and \(10\)--\(90\%\) empirical-quantile ranges of the non-degenerate
\(\mu\) values. For each benchmark, we quantify this dependence through
\begin{equation}
\Delta_{\rm fit}
=
\max_q \widehat d_{\rm TwoNN}^{(q)}
-
\min_q \widehat d_{\rm TwoNN}^{(q)},
\end{equation}
where \(q\) runs over the three fitted quantile ranges. The largest value of
\(\Delta_{\rm fit}\) among the four benchmarks is \(0.27\). This variation is
used as a fitting-convention check rather than as an independent statistical
uncertainty.

As a diagnostic of lattice effects, we also consider the \(L_\infty\)
distance,
\begin{equation}
d_{L_\infty}(\mathbf{x},\mathbf{y})
=
\max_a |x_a-y_a|.
\end{equation}
Because this metric depends only on the largest coordinate difference, many
distinct points in the integer exponent lattice are assigned the same
distance. Consequently, for \(67.4\)--\(71.4\%\) of the textures, the first
and second nearest neighbours are distinct but lie at the same positive
distance, \(r_{2,i}=r_{1,i}>0\). Their TwoNN ratio is therefore
\(\mu_i=1\), so that \(\log\mu_i=0\), and they provide no information about
the fitted TwoNN slope. Since this degeneracy affects most of the sample, the
\(L_\infty\) calculation cannot provide a reliable alternative estimate of
the intrinsic dimension. It serves instead to demonstrate that the choice of metric can make the
discrete lattice structure of the exponent space dominate the
nearest-neighbour statistics.

Finally, we examine whether the TwoNN result depends strongly on the number of
textures included in the analysis. For each full \(N=1000\) benchmark, we
generate 100 random subsets containing
\begin{equation}
N=300,\qquad 500,\qquad 800
\end{equation}
textures and recompute \(\widehat d_{\rm TwoNN}\) independently for every
subset. The subsets are drawn without replacement, meaning that a texture
cannot occur more than once within the same subset. This choice is important
because ordinary resampling with replacement would create artificial
duplicates and hence zero nearest-neighbour distances. Different random
subsets may nevertheless contain some of the same textures.
For each benchmark and sample size, the resulting distribution of 100 estimates
is summarized by its median and by the central \(68\%\) range,
\begin{equation}
\left[
Q_{0.16}\!\left(\widehat d_{\rm TwoNN}\right),
Q_{0.84}\!\left(\widehat d_{\rm TwoNN}\right)
\right],
\end{equation}
where \(Q_{0.16}\) and \(Q_{0.84}\) denote the 16th and 84th percentiles \footnote{This
range is used as a descriptive measure of the variation among random subsets
and is not interpreted as a confidence interval for a unique manifold
dimension.}
The median estimates change only moderately as the sample size is reduced,
while the percentile ranges become wider at smaller \(N\), as expected. The
largest variation occurs for the full-CKM benchmark at \(N=300\), for which
\begin{equation}
\operatorname{median}\!
\left(\widehat d_{\rm TwoNN}\right)=10.45,
\qquad
Q_{0.16}\text{--}Q_{0.84}=9.51\text{--}11.08.
\end{equation}
The value \(9.51\) is also the smallest 16th percentile found among all four
benchmarks and all tested sample sizes. Thus, even when only 300 of the 1000
available textures are retained, the sampling fluctuations do not move the
TwoNN estimates into a very-low-dimensional regime.

Taken together, these checks show that the qualitative conclusion of high
effective dimensionality is not driven by duplicate textures, the modest
number of Euclidean distance ties, the selected fitting interval, or the use
of the complete \(N=1000\) samples. At the same time, the strong degeneracy
observed for \(L_\infty\) and the moderate shifts induced by coordinate
rescaling confirm that the numerical results are metric-dependent diagnostics
of a finite, discrete point cloud. They should therefore not be interpreted
as precise measurements of a unique underlying manifold dimension.

\section{Conclusions}
\label{sec:conclusions}

In this work we have performed an exploratory study of the geometry of viable
Froggatt--Nielsen-like flavor textures in the raw exponent coordinates. Each texture
was represented as a point in the 18-dimensional space of up- and down-type FN
exponents, and viability was imposed through independent multiplicative cuts on
quark mass ratios and CKM observables. By comparing four benchmark ensembles,
from masses-only constraints to the inclusion of all the CKM moduli considered
in this analysis, we investigated how increasingly restrictive flavor data
shape the geometry of the accepted landscape.

The main result is that the viable FN textures do not exhibit strong
compression to a very-low-dimensional structure in the raw exponent space.
Principal-component analysis shows that the accepted point clouds are not
concentrated near a low-dimensional linear subspace: in all four benchmark
datasets, \(N_{90}=15\) principal components are required to account for
90\% of the total variance. The inclusion of CKM constraints enhances the
leading principal components and makes the accepted point cloud more
anisotropic, but it does not reduce \(N_{90}\). Mixing observables therefore
introduce additional geometric organization into the viable landscape, but do
not collapse it onto a low-dimensional linear subspace.

Nearest-neighbour intrinsic-dimensionality diagnostics support the same
qualitative picture. Under the primary raw-\(L_2\) convention, the TwoNN
estimates are of order \(10\)--\(12\), while the Levina--Bickel cross-check
gives similarly high effective dimensionalities over the neighbourhood scales
considered. These values should not be interpreted as precise measurements of
a unique underlying manifold dimension. The integer-valued nature of the
exponent space, the sensitivity to the distance metric, and the finite number
of sampled textures all limit such an interpretation. Nevertheless, the
robustness checks show that the qualitative result is not driven by duplicate
textures, the modest number of Euclidean distance ties, the selected TwoNN
fitting interval, or the use of the complete \(N=1000\) samples. We therefore
find no evidence that the accepted FN textures form a very-low-dimensional
point cloud in the raw 18-dimensional exponent coordinates.

This negative result is informative, but should be interpreted within the
coordinate representation and metrics considered here. It does not imply that
flavor theory space lacks hidden geometric structure. Rather, it shows that
such a structure is not manifested as strong low-dimensional compression in
the raw FN exponent representation. One possible interpretation is that the
FN exponents are not the coordinates best adapted to the physically relevant
geometry. Establishing this possibility, however, requires the construction
and comparison of alternative representations. The present study should therefore be viewed as a first step toward a broader
geometric approach to the flavor problem. 

Future work may investigate
observable-space geometry, approximate fibers of the map from textures to
observables, and metrics defined by the sensitivity of physical observables to
texture deformations. These approaches may determine whether the apparent
redundancy of flavor model building is organized by a geometric structure that
is not manifest in the raw exponent coordinates.
\section*{Acknowledgements}
D.M. is deeply indebted to Alessio Giarnetti for  reading the manuscript and providing important suggestions. 
\bibliographystyle{unsrt}  
\bibliography{bibliography}

\end{document}